\documentclass[twocolumn,twocolappendix]{aastex6}
\usepackage{amsmath,amstext}
\usepackage[version=4]{mhchem}
\usepackage{ragged2e}
\usepackage{apjfonts}
\usepackage{gensymb}
\usepackage{multirow}
\usepackage{tikz}
\usepackage{graphics}
\usepackage[figure,figure*]{hypcap}
\usepackage[para,online,flushleft]{threeparttable}
\usepackage{comment}
\usepackage{tablefootnote}
\usepackage{subscript}

\makeatletter
\newcounter{reaction}
\renewcommand\thereaction{R\arabic{reaction}}
\@addtoreset{reaction}{chapter}
\newcommand\reactiontag%
{\refstepcounter{reaction}\tag{\thereaction}}
\newcommand\reaction@[2][]%
{\begin{equation}\ce{#2}%
\ifx\@empty#1\@empty\else\label{#1}\fi%
\reactiontag\end{equation}}
\newcommand\reaction@nonumber[1]%
{\begin{equation*}\ce{#1}\end{equation*}}
\newcommand\reaction%
{\@ifstar{\reaction@nonumber}{\reaction@}}
\makeatother

\shorttitle{Oxygen in Non-1 bar Atmospheres}
\shortauthors{James and Hu}

\begin{document}

\title{Photochemical Oxygen in Non-1 Bar CO$_2$ Atmospheres of Terrestrial Exoplanets}

\author{Tre'Shunda James\altaffilmark{1,2}}\author{ Renyu Hu\altaffilmark{1,3}}

\altaffiltext{1}{Jet Propulsion Laboratory, California Institute of Technology, Pasadena, CA 91109, USA; renyu.hu@gmail.com}
\altaffiltext{2}{Occidental College, Los Angeles, CA 90041, USA}
\altaffiltext{3}{Division of Geological and Planetary Sciences, California Institute of Technology, Pasadena, CA 91125, USA}

\begin{abstract}

Atmospheric chemistry models have shown molecular oxygen can build up in CO$_2$-dominated atmospheres on potentially habitable exoplanets without input of life. Existing models typically assume a surface pressure of 1 bar. Here we present model scenarios of CO$_2$-dominated atmospheres with the surface pressure ranging from 0.1 to 10 bars, while keeping the surface temperature at 288 K. We use a one-dimensional photochemistry model to calculate the abundance of O$_2$ and other key species, for outgassing rates ranging from a Venus-like volcanic activity up to $20\times$ Earth-like activity. The model maintains the redox balance of the atmosphere and the ocean, and includes the pressure dependency of outgassing on the surface pressure. Our calculations show that the surface pressure is a controlling parameter in the photochemical stability and oxygen buildup of CO$_2$-dominated atmospheres. The mixing ratio of \ce{O2} monotonically decreases as the surface pressure increases at the very high outgassing rates, whereas it increases as the surface pressure increases at the lower-than-Earth outgassing rates. Abiotic \ce{O2} can only build up to the detectable level, defined as $10^{-3}$ in volume mixing ratio, in 10-bar atmospheres with the Venus-like volcanic activity rate and the reduced outgassing rate of \ce{H2} due to the high surface pressure. Our results support the search for biological activities and habitability via atmospheric O$_2$ on terrestrial planets in the habitable zone of Sun-like stars.


\end{abstract}


\section{Introduction}

Recent discoveries of exoplanets (i.e., planets orbiting stars other than the Sun) have consistently sparked new research interest. Whether or not these planets are habitable and under what conditions allow them to be has been driving exoplanet research. Today, scientists rely on remote sensing to detect and study exoplanets. The {\it Kepler} mission has confirmed over 2,000 exoplanets; more than 20 of them are less than twice Earth-size and receive the right amount of stellar irradiation that would supports a liquid-water ocean on their surface (based on the NASA Exoplanet Archive). If rocky, they are candidates for habitable exoplanets. The Transiting Exoplanet Survey Satellite (\textit{TESS}) has been launched in 2018, and will cover an area of sky 400$\times$ larger than {\it Kepler} and find many more potentially habitable worlds \citep{Sullivan2015}. The James Webb Space Telescope (\textit{JWST}) will have the capability to perform infrared spectroscopy of exoplanet atmospheres, including some of the potentially habitable worlds found by ground-based transit surveys \citep{Dittmann2017, Gillon2017} and \textit{TESS} \citep{Schwieterman2016}. Therefore, the search for signatures of habitability and bio-activities from exoplanet spectra is eminent.


Oxygen is considered a primary biosignature gas, thanks to its predominantly biogenic origins on Earth and its spectral features in the near-infrared \citep[e.g.,][]{1965Nature....207...568L,1967Icarus....7...149H, Sagan1993}.
Abiotic processes such as photodissociation of H$_2$O and CO$_2$ can also produce oxygen, and if accumulated in the atmosphere, the abiotic oxygen would impede the use of oxygen as a biosignature gas \cite[see][and references therein]{meaddows2017}. In the case of an N$_2$-dominated atmosphere, it has been shown that if the partial pressure of N$_2$ is too small, water vapor cannot be trapped in the troposphere and the upper part of the atmosphere would be wet \citep{wordsworth2014}. This leads to enhanced photodissociation of water and buildup of abiotic oxygen. In the case of a CO$_2$-dominated atmosphere, early investigations suggest massive buildup of oxygen \citep{Selsis2002}; the result was later called into question based on the redox balance in the atmosphere \citep{Segura2007}. Recent models that explicitly balance the redox budget of the atmosphere suggest that abiotic oxygen can sometimes accumulate to a detectable level, if the volcanic emission rate is very low \citep{2012ApJ....761...166H, 2014ApJ....792...90D} or the hydrogen content of the atmosphere is low \citep{Gao2015}. If the planet is around an M dwarf star, the accumulation of O$_2$ will be much larger. Even in the case of volcanic emission rates as high as Earth, abiotic O$_2$ can still accumulate to the detectable level \citep{Tian2014,Harman2015}. In this work, we define the detectable level as $\sim10^{-3}$ in volume mixing ratio, as suggested by evaluating direct detection of the O$_2$ A band in space \citep{DesMarais2002,Segura2003}. One may also consider a potential-false-positive level to be $10^{-3}$ the present atmospheric level (PAL) or $2\times10^{-4}$, the oxygen level of Earth during most of the Proterozoic Eon with oxygenic photosynthesis \citep{Planavsky2014}. This level of oxygen is not detectable via direct imaging in the near infrared, but may be detectable in the far future via \ce{O3} features in the thermal infrared \citep{Leger1993,Segura2003} and ultraviolet wavelengths \citep{2014ApJ....792...90D}.


Previous work in analyzing abiotic accumulation of oxygen in CO$_2$-dominated atmospheres assume the surface pressure to be $\sim1$ bar. In reality, there is no reason to assume {\it a priori} the size of a habitable exoplanet's atmosphere. First, small exoplanets have diverse compositions. For example, the TRAPPIST-1 planets, with radii of approximately Earth's radius, have very different bulk densities \citep{Grimm2018} and therefore, some of the planets may have massive atmospheres and some may not. Second, the very concept of the ``habitable zone'' implies that the mass of atmospheric CO$_2$ increases as the stellar irradiation decreases to maintain the conditions for liquid water oceans at the surface \citep{Kasting:1993zz, Bean2017}. Atmospheric circulation models have shown that TRAPPIST-1 e would be habitable if it has a 1-bar CO$_2$-dominated atmosphere, and TRAPPIST-1 f would be habitable if it has a 2-bar CO$_2$ atmosphere \citep{Wolf2017, Turbet2017}. In all, the size of a habitable exoplanet's atmosphere is unknown and must not be simply assumed as 1 bar.


Here we study abiotic production and accumulation of oxygen in CO$_2$-dominated atmospheres with varied surface pressures, using a photochemistry model. Our goal is to explore the effect that changing the surface pressure can have upon the abundance of O\textsubscript{2} in the atmosphere. By investigating the abiotic production of oxygen in non-1 bar atmospheres, we further clarify the conditions that the false-positive scenarios would occur. This paper is organized as follows. Section 2 describes the photochemistry model and the specifics of the simulated atmospheres. In Section 3, we present model results with the surface pressure ranging from 0.1 to 10 bars. We discuss their implications in Section 4 and conclude in Section 5.

\section{Methods}


\begin{table*}[ht]

\caption{Parameters for Modeled Terrestrial Exoplanets}
\centering
\begin{tabular}{lccccc}
\hline \hline

Parameters &   \multicolumn{5}{c}{{CO$_2$}-dominated Atmosphere}\\
\hline
Main Component & \multicolumn{5}{c}{90\% CO\textsubscript{2} ,  10\%  N\textsubscript{2}} \\
Mean Molecular Mass & \multicolumn{5}{c}{42.4}\\
\hline
\textit{Surface  Pressure} & \multicolumn{5}{c}{0.1 bar\hspace{1.1cm}0.3 bar \hspace{1cm}1 bar \hspace{1.1cm}3 bars\hspace{1.1cm}10 bars}\\
\hline
\textit{Planetary Parameters}& \\
Stellar Type &  \multicolumn{5}{c}{G2V}\\
Mass &  \multicolumn{5}{c}{1 M\textsubscript{$\oplus$}}\\
Radius &  \multicolumn{5}{c}{1 R\textsubscript{$\oplus
$}}\\

Semi-major Axis  &\multicolumn{5}{c}{1.1 AU \hspace{.9cm} 1.2 AU \hspace{.9cm}1.3 AU \hspace{.9cm}1.5 AU\hspace{.9cm} 2.1 AU}\\
\hline
\textit{Temperature-Pressure Profile}\\
Surface Temperature&  \multicolumn{5}{c}{288 K }\\
Tropopause Altitude&  \multicolumn{5}{c}{9.77 km}\\
Temperature Above Tropopause & \multicolumn{5}{c}{175 K }\\
Top-of-Atmosphere (0.1 Pa) Altitude& \multicolumn{5}{c}{43 km\hspace{1cm} 47 km\hspace{1cm} 52 km \hspace{1cm}55 km\hspace{1cm} 59 km}\\
\hline
\textit{Water and Rainout}\textsuperscript{a,b}\\
Liquid Water Ocean  & \multicolumn{5}{c}{Yes}\\
Water Vapor Boundary Condition& \multicolumn{5}{c}{\textit{f}(H\textsubscript{2}O) = 0.01}\\
Rainout rate& \multicolumn{5}{c}{Earth-like}\\

\hline
\hline
\textit{Gas Emission\textsuperscript{c}} (molecule cm$^{-2}$ s$^{-1}$) &\multicolumn{5}{c}{\textit{Very High Emission}  \hspace{.9cm}\textit{High Emission} \hspace{.9cm}\textit{Low Emission} \hspace{.9cm} \textit{Very Low Emission}}\\
CO & \multicolumn{5}{c}{ $3\times 10^{10}$\hspace{2cm} $1.5\times 10^{9}$\hspace{1.6cm} $2.0\times 10^8$\hspace{1.8cm} $2.5\times 10^7$}\\
H\textsubscript{2} & \multicolumn{5}{c}{ $3\times 10^{10}$, d\hspace{2cm} $1.5\times 10^{9}$\hspace{1.6cm} $2.0\times 10^8$\hspace{1.8cm} $2.5\times 10^7$}\\
SO\textsubscript{2}& \multicolumn{5}{c}{ $3\times 10^{10}$, d\hspace{2cm} $1.5\times 10^{9}$\hspace{1.6cm} $2.0\times 10^8$\hspace{1.8cm} $2.5\times 10^7$}\\
H\textsubscript{2}S & \multicolumn{5}{c}{ $3\times 10^{9}$\hspace{2.1cm} $1.5\times 10^{8}$\hspace{1.6cm} $2.0\times 10^7$\hspace{1.8cm} $2.5\times 10^6$}\\
\hline

\RaggedLeft
\textbf{Notes.}\\
\multicolumn{6}{l}{\textsuperscript{a}
The rainout rate of the non-soluble species CO, CH$_4$, H$_2$, O$_2$ and C$_2$H$_6$ are generally zero.}\\

\multicolumn{6}{l}{\textsuperscript{b}
The deposition velocities of gases follow the ones used in \cite{2012ApJ....761...166H}.}\\

\multicolumn{6}{l}{\textsuperscript{c}
The volcanic gas emission rates are assigned as lower boundary conditions in each scenario of emission. H$_2$O and CO$_2$ are}\\
\multicolumn{6}{l}{ considered to be abundant in the system.}\\

\multicolumn{6}{l}{\textsuperscript{d}
Assumed to be $3\times10^9$ molecule cm$^{-2}$ s$^{-1}$ in the special case where we include the effect of surface pressure on the outgassing speciation.}\\


\end{tabular}

\label{Tab:1}

\end{table*}

\subsection{Photochemistry Model}

The photochemistry model we utilize has previously been used to study abiotic oxygen in reducing, weakly oxidizing and highly oxidizing atmospheres \citep{2012ApJ....761...166H, Hu2013}, as well as biosignatures in Super-Earths with H\textsubscript{2}-rich atmospheres \citep{2013ApJ....777...95S}. Using the known atmospheric pressure-temperature profile and background composition, the code is able to accurately predict the amount of trace gases in both Earth's and Mars' atmospheres. The code finds steady state solutions of species at each altitude of the model atmosphere. Tracing the kinetics of hundreds of chemical reactions, the model can compute the concentrations of 111 molecules and aerosols. A complete list of molecules and reactions can be found in  \cite{2012ApJ....761...166H}. The model uses a self-adjusting time step to track the progress of balancing products and loss from all chemical and photochemical reactions towards reaching a steady state. As the code gets closer to convergence, the time step increases. The code is considered as converged once the time step becomes very large, e.g., $10^{17}$ s.


\subsection{CO\textsubscript{2}-Dominated Atmospheres}

\cite{2012ApJ....761...166H} present benchmark cases with 1-bar atmospheres on terrestrial exoplanets ranging from H$_2$-dominated to CO$_2$-dominated. Here we focus on the CO$_2$-dominated atmosphere scenarios and expand to include surface pressures ranging from 0.1 to 10 bars. Following the benchmark case of \cite{2012ApJ....761...166H}, we consider the background composition of 90\% CO\textsubscript{2} and 10\% N\textsubscript{2}, and include 57 C-, H-, O-, N-, and S-bearing species in the model. We focus on the planets around Sun-like stars in this work; the stability of non-1-bar CO$_2$ atmospheres around M stars will be addressed in a separate paper. We do not include the potential effects of lightning \citep{Rimmer2016,Hodosan2016,Wong2017} in this study. Specific parameters for our model atmospheres are tabulated in Table \ref{Tab:1}.



\textit{Temperature-Pressure Profile} To model habitable exoplanets, we adopt a surface temperature of 288 K. The temperature profile is then set to follow an appropriate adiabatic lapse rate (i.e., the convective layer) until 175 K and is assumed to be constant above (i.e., the radiative layer). We simulate the atmospheres up to an altitude that corresponds to 0.1 Pa. The temperature profile is consistent with significant greenhouse effects in the convective layer and no additional heating above the convective layer for habitable exoplanets. We caution that the latter point may not be entirely valid because appreciable amounts of ozone are produced in some of our scenarios,  which can lead to heating above the convective layer.

\textit{Semi-major Axis.} To preserve a surface temperature of 288 K as the surface pressure changes, we adjust the orbital distance. We assume a circular orbit and adjust the semi-major axis by balancing the incoming stellar irradiation and the outgoing thermal irradiation for full heat redistribution.
The semi-major axis ($a$) is estimated by
\begin{equation}
a = \sqrt{\frac{L(1-A_B)}{16\pi I}},
\end{equation}
where $L$ is the stellar luminosity, $A_B$ is the Bond albedo, and $I$ is the thermal emission irradiation flux. We assume the terrestrial value for $A_B$ and use the pre-calculated formula for $I$ as a function of the CO$_2$ partial pressure and the surface temperature \citep{WILLIAMS1997254}. The range of surface pressure we consider is within the range of validity of the formula.




\begin{figure*}[t]
\includegraphics[width=\linewidth]{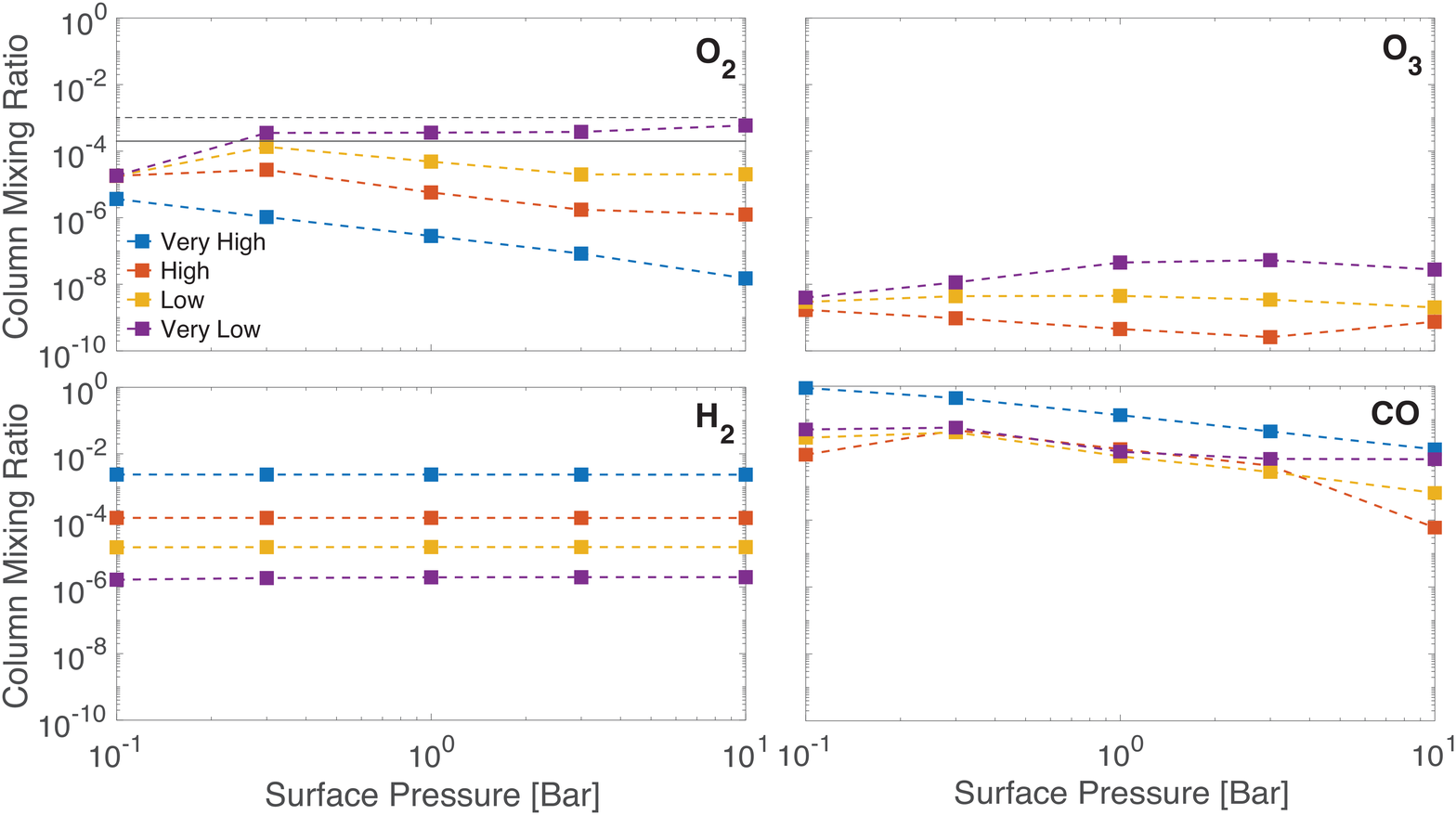}
\caption{Mixing ratios of the key species in the CO$_2$-dominated atmosphere as a function of the surface pressure. Four surface gas emission scenarios defined in Table \ref{Tab:1} are shown in four different colors. The black dashed line in the O$_2$ panel shows the level above which the abiotic O$_2$ would be detectable ($10^{-3}$), and the black solid line shows the oxygen level of Proterozoic Earth ($2\times10^{-4}$). Each point is summarized from a converged simulation of the full photochemical model. The abundance of O$_2$ increases with the surface pressure at very low emission rates, whereas it decreases with the surface pressure at very high emission rates.}
\label{Fig: Figure 1}
\end{figure*}

\begin{figure}[t]
\centering
\includegraphics[width=\linewidth]{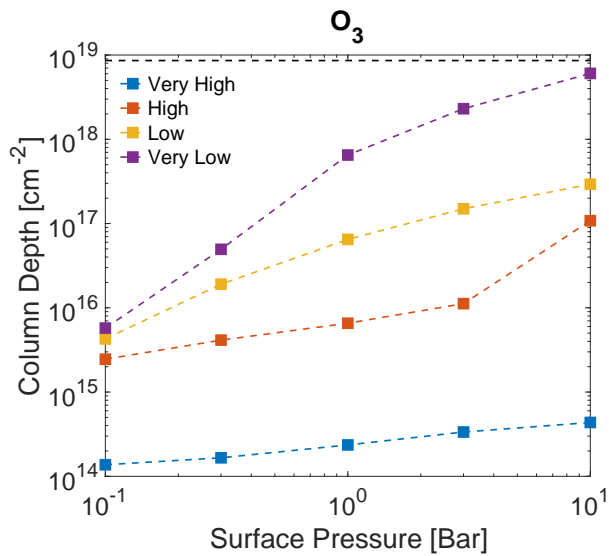}
\caption{Column depth of \ce{O3} as a function of the surface pressure and the surface emission scenarios. The black dashed line shows the column depth of \ce{O3} of present-day Earth for comparison.}
\label{Fig: Figure 1X}
\end{figure}

\begin{figure*}[t]
\includegraphics[width=\linewidth]{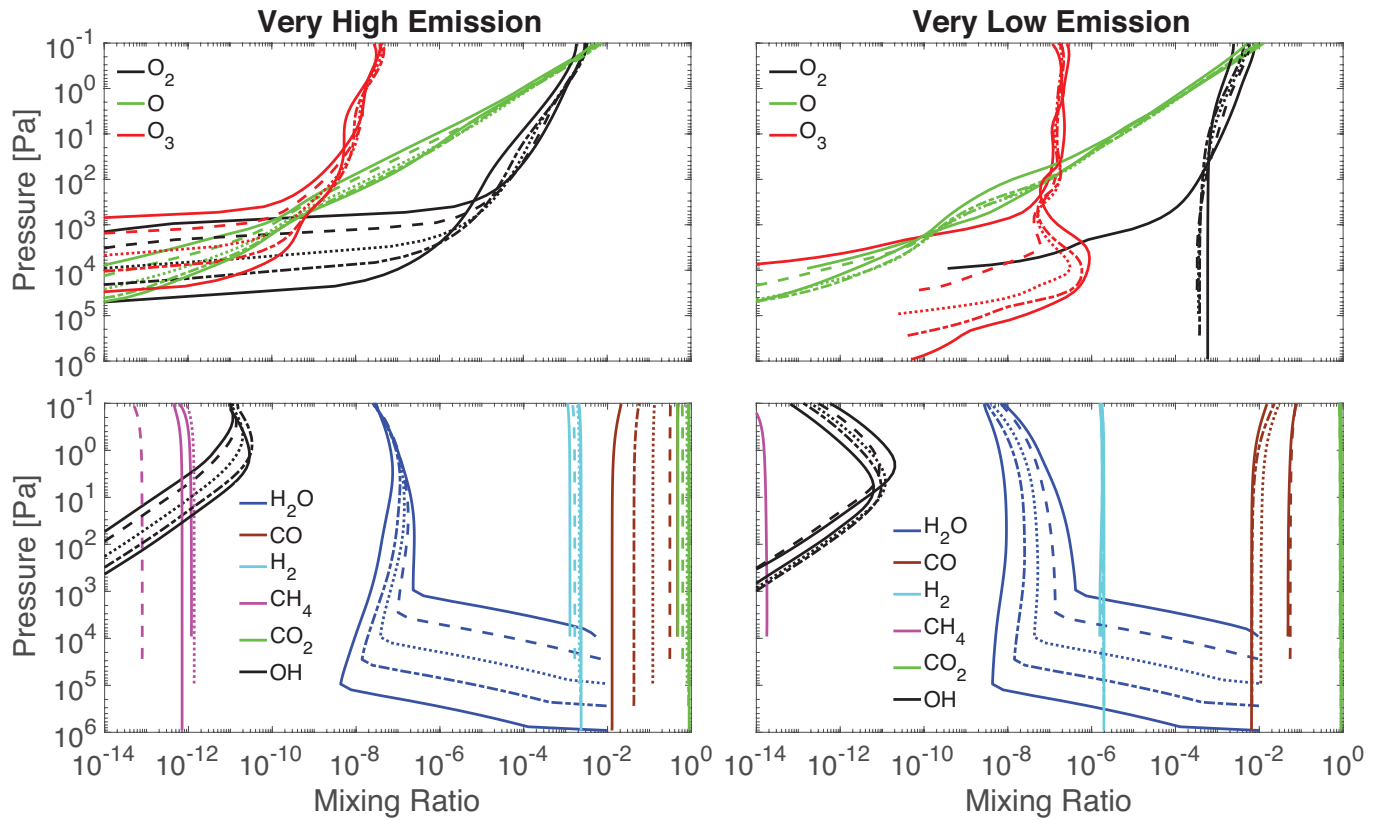}
\caption{Comparison between the very high emission scenario and the very low emission scenario in terms of the mixing ratio profiles of the key species. Solid lines are used for both 0.1 bar and 10 bar cases, and dashed, dotted, and dash-dotted lines correspond to 0.3, 1, and 3 bar cases, respectively. At the very high emission rate, no O$_2$ is accumulated at the surface regardless of the surface pressure, whereas at the very low emission rate, O$_2$ can accumulate when the surface pressure is $>0.1$ bar.
}
\label{Fig: Figure 2}
\end{figure*}

\begin{figure*}[t]
\centering
\includegraphics[width=\linewidth]{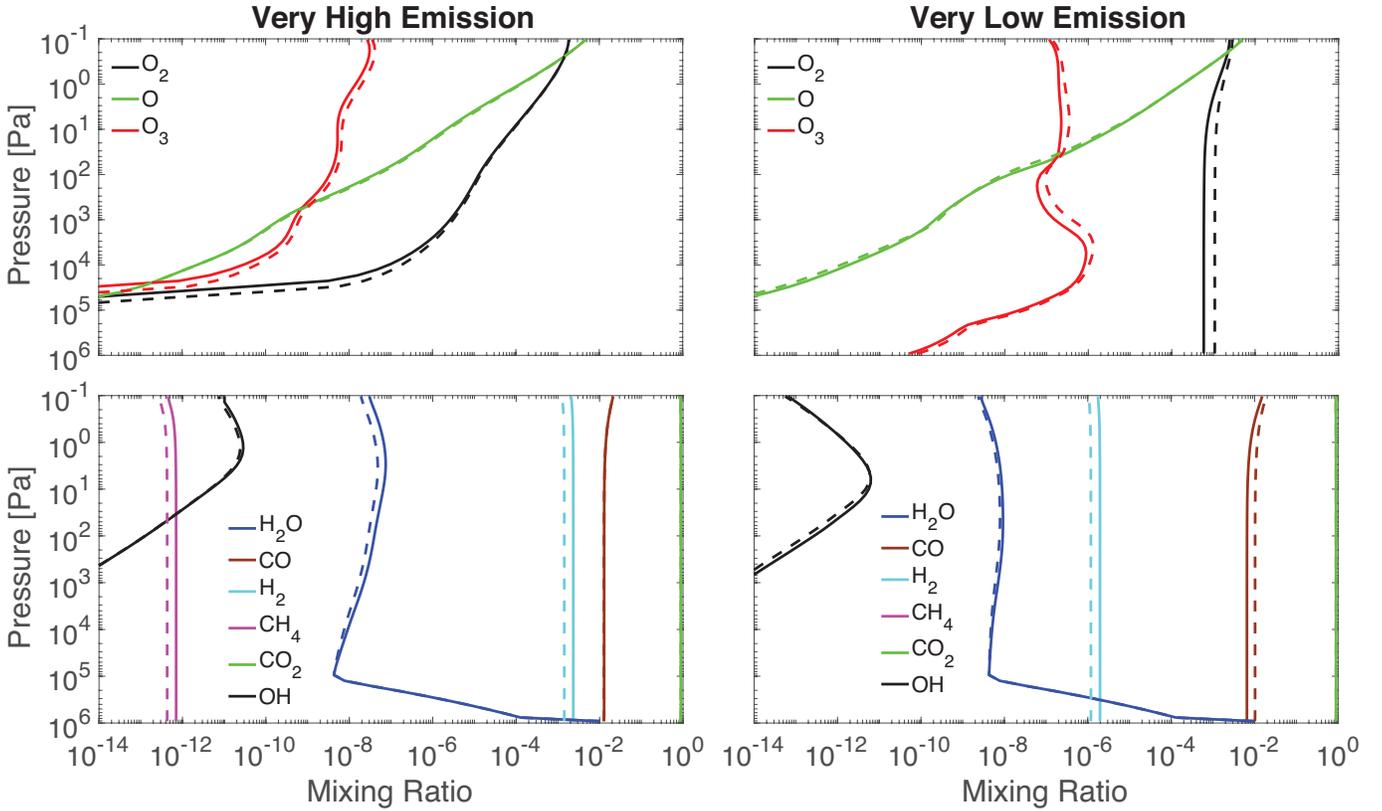}
\caption{Comparison of two speciation models for the 10 bar surface pressure atmosphere. The solid lines correspond to the outgassing rates tabulated in Table \ref{Tab:1}, and the dashed lines include 10-fold less \ce{H2} and \ce{SO2} due to the effect of surface pressure to the outgassing speciation \citep{GAILLARD2014307}. No significant difference is found between the two speciation models at very high emission, but the abundance of \ce{O2} increases by 80\% for the 10-fold less \ce{H2} and \ce{SO2} at very low emission.}
\label{Fig: Figure 2x}
\end{figure*}

\begin{figure*}[t]
\includegraphics[width=\linewidth]{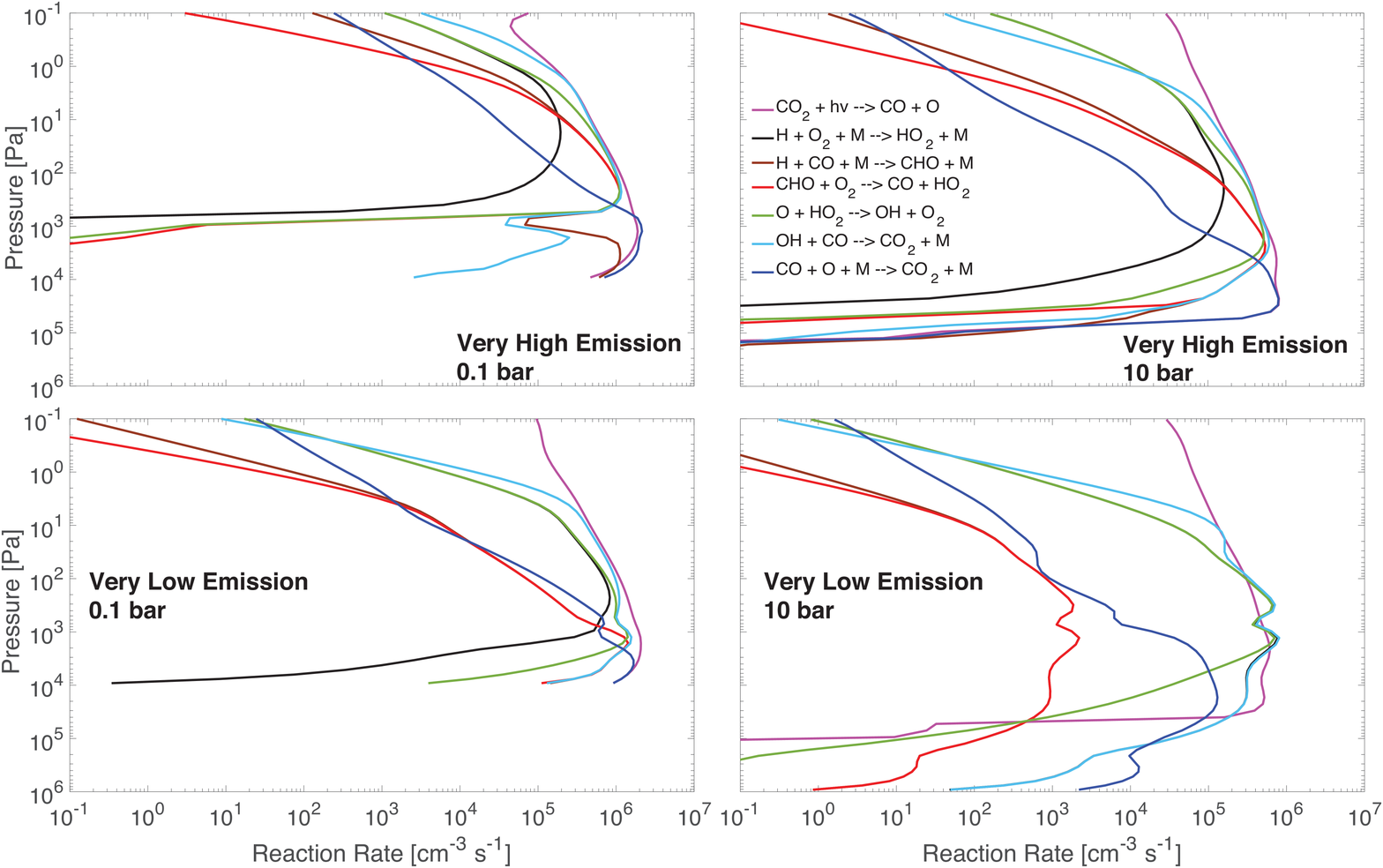}
\caption{Key chemical reactions that contribute to the combination of CO and O. For all cases the direct combination of CO + O + M dominates the lower atmosphere and the reaction \ce{OH + CO -> CO2 + H} dominates the middle atmosphere. At the very high emission rate, the cycle of CHO (Reactions \ref{r:5} and \ref{r:6}) dominates the production of HO$_2$; whereas at the very low emission rate, direct combination of H and O$_2$ (Reaction \ref{r:3}) dominates the production of HO$_2$.}
\label{Fig: Figure 3}
\end{figure*}

\subsection{Boundary Conditions}

\textit{Outgassing.} We simulate scenarios of various surface gas emission rate to explore its effect on the atmospheric abundance of abiotic oxygen.
The gas emission rate ($\phi$) is assumed to be dominated by volcanic outgassing, and is parameterized by
\begin{equation}
\phi= V \times \varphi ,
\end{equation}
where $V$ is the magma production rate and $\varphi$ is the volatile content released by unit volume of magma. For the magma production rate, we explore a wide range. We consider the volcanic production rates of the present-day Earth (30 km$^3$ yr$^{-1}$, \cite[e.g.,][]{GAILLARD2014307}) and the present-day Venus (0.5-3 km$^3$ yr$^{-1}$, \cite{JGRE:JGRE20250}), and they are defined to be the ``high emission'' and the ``very low emission'' scenarios, respectively. We use the lower end of estimate of \cite{JGRE:JGRE20250} to represent an endmember scenario of very low emission. For completeness, we also consider a ``low emission'' scenario where the volcanic production rate is the geometric mean of the high emission and very low emission scenarios (3.9 km$^3$ yr$^{-1}$), as well as a ``very high emission'' scenario to have a volcanic production rate $\sim20\times$ that of the present-day Earth (600 km$^3$ yr$^{-1}$). Such a high rate may be found in younger-than-Earth planets or planets that receive large tidal dissipation.

The volcanic production rate of 0.5 km$^3$ yr$^{-1}$ can be converted to $4.3\times10^4$ kg s$^{-1}$, for a density of 2700 kg m$^{-3}$ (typical for mid-ocean ridge basalts). For a volcanic gas content of approximately 1000 ppm in mass and a mean molecular weight of 20 \citep[e.g.,][]{GAILLARD2014307}, this corresponds to $2.1\times10^3$ mol s$^{-1}$. Averaging over an Earth-radius planet, this is $\sim2.5\times10^8$ molecules cm$^{-2}$ s$^{-1}$.


For speciation, we follow the planetary magmatic degassing calculations of \cite{GAILLARD2014307}. Consider typical mid-ocean ridge basalts degassing at a surface pressure ranging from 0.1 to 3 bars: H$_2$, CO, and SO$_2$ have mole fractions of $\sim0.1$ in the volcanic gas (Figure 4 of \cite{GAILLARD2014307}), and at 10 bars, the mole fractions of H$_2$ and SO$_2$ reduce to $\sim0.01$. The mole fractions of the elemental sulfur and H$_2$S are $\sim0.01$ for all surface pressures. In this study, to isolate the effects of changing surface pressure, we adopt in the standard model a mole fraction of 0.1 for H$_2$, CO, and SO$_2$ and 0.01 for H$_2$S. This is applied to all surface pressures. We additionally study a special case for the 10 bar surface pressure atmosphere, where we take into account less degassing of H$_2$ and SO$_2$ under high surface pressures by assuming a mole fraction of 0.1 for CO, and 0.01 for H$_2$, SO$_2$, and H$_2$S. According to our estimate, the emission rate of H$_2$ of an abiotic Earth is $1.5\times10^9$ molecules cm$^{-2}$ s$^{-1}$, consistent with \cite{Sleep2007}. The gas emission rates we model are tabulated in Table \ref{Tab:1}.


In addition to the gas emission, we include diffusion-limited escape for H and H$_2$ at the top of the atmosphere in the same way as \cite{2012ApJ....761...166H}.



\textit{Deposition and Redox Balance.} We include dry deposition and rainout of species from the atmosphere in the same way as \cite{2012ApJ....761...166H}. In particular, the deposition velocity of CO to the ocean is assumed to be $10^{-8}$ cm s$^{-1}$, and that of \ce{O2} is assumed to be zero. This assumption considers that CO is slowly converted to acetate and precipitate in the ocean \citep{Kharecha2005}, and ignores various potential sinks for \ce{O2}. \cite{Harman2015} considered recombination of CO and \ce{O2} in hydrothermal systems, oxidation of formate in solution, and oxidation of ferrous iron as the potential sinks of \ce{O2} in the ocean, and suggested that these sinks should be minor if using the Earth's process rates as the guide. We have tested the model by changing the deposition velocity of \ce{O2} to $10^{-8}$ cm s$^{-1}$, the same as that of CO. The resulting mixing ratio profiles are not distinguishable with the standard assumption.
If rapid processes exist in the ocean to directly recombine CO and \ce{O2}, their deposition velocities would be on the order of $10^{-4}$ cm s$^{-1}$ \citep{2014ApJ....792...90D, Harman2015}; and this would greatly reduce the amount of CO and \ce{O2} in the atmosphere in the steady state. Since there is no such processes known on Earth, we do not include them in the model.

We enforce redox balance in the atmosphere and in the ocean for all simulated scenarios. The redox balance says that the total redox influx to system (i.e., surface emission) should be balanced by the total redox outflux from the system, otherwise the system is being oxidized or reduced. The redox balance is equivalent to the conservation of the total number of electrons in the system. When applied to the atmosphere, the balance is strictly enforced by the mass balance and the convergence of the photochemical model itself. Specifically, we solve all molecules in the mass conservation equation, without assuming any ``fast'' species.
It follows that the imbalance between the surface emission and the escape of hydrogen, if any, does not cause the redox of the atmosphere to change over time (it must be balanced by dry deposition and rainout), or,
\begin{equation}
\Phi(\rm{Outgassing}) = -\Phi(\rm{Escape}) - \Phi(\rm{Deposition}),
\end{equation}
where $\Phi$ is the flux of net reducing species into the atmosphere, following the definition in \cite{2012ApJ....761...166H}.

Recently, it has been realized that $\Phi(\rm{Deposition})$ may not be arbitrary in realistic planetary scenarios, as it represents a net transfer of reducing or oxidizing species from the atmosphere to the ocean \citep{2014ApJ....792...90D, Harman2015}. The key idea is that {\it there shall be no net reducing species into the ocean on a planet without life.} This is based on an analog to Earth as a terrestrial planet, where the only ways that the ocean can remove reducing species are burial of organic matter and burial of sulfide, both involving life \cite[e.g.,][]{Harman2015}. If this is universally applicable to exoplanets, we must impose
\begin{equation}
\Phi(\rm{Deposition}) \geq 0. \label{eq.redox}
\end{equation}
Moreover, if $\Phi(\rm{Deposition}) >0 $, net oxidizing species is deposited into the ocean. While this is possible geologically (e.g., via burial of magnetite), this would reduce the abundance of O$_2$ in the atmosphere. To simulate the limiting case for oxygen buildup in the atmosphere, we impose
\begin{equation}
\Phi(\rm{Deposition}) = 0. \label{eq.redox0}
\end{equation}

Equation (\ref{eq.redox0}) is essentially equivalent to $\Phi(\rm{Outgassing}) + \Phi(\rm{Escape})=0$, and is not automatically enforced by the photochemical model. We enforce this condition the same way as \cite{Harman2015}, in that we include a return flux of hydrogen from the ocean to the atmosphere, if finding $\Phi(\rm{Deposition}) \neq 0$ for a converged solution. We then relaunch the simulation and repeat the process until the condition in Equation (\ref{eq.redox0}) is sufficiently satisfied. In practice we require an imbalance of the global redox budget to be no larger than 1\% of the outgassing redox flux (Table \ref{Tab2}). This way, our converged models satisfy redox balance for both the atmosphere and the ocean.

\section{Results} \label{section:results}

We find that the trend of photochemically produced \ce{O2} in the CO$_2$-dominated atmosphere is different between the high and the low emission scenarios: when the emission rate is high (e.g., $20\times$ Earth's volcanic activity level, a.k.a. the very high emission scenario in this study), the mixing ratio of O$_2$ decreases as a function of the surface pressure; and when the emission rate is low (e.g., Venus's volcanic activity level, a.k.a. the very low emission scenarios in this study), the mixing ratio of O$_2$ increases as a function of the surface pressure (Figure \ref{Fig: Figure 1}). In between the two endmembers, the Earth-like emission rates (a.k.a. the low and high emission scenarios in this study) have the O$_2$ mixing ratio peaks at $\sim0.3$ bar to $10^{-4}$ (Figure \ref{Fig: Figure 1}). In addition, the column depth of \ce{O3} photochemically produced in the atmosphere increases with the surface pressure, and also is higher for lower surface emission rates (Figure \ref{Fig: Figure 1X}).

The redox balance of the atmosphere and the ocean is maintained for all scenarios. Table \ref{Tab2} shows the redox balance for the four endmember scenarios of the very high and very low emission rate and the 0.1 bar and 10 bar surface pressure. We see that for all models the escape flux balances the outgassing flux. Figure \ref{Fig: Figure 1} shows that the mixing ratio of \ce{H2} depends on the outgassing flux but not on the surface pressure, which is a direct consequence of maintaining the balance between hydrogen escape and outgassing of reducing species. Also, the net deposition flux is balanced by the return \ce{H2} flux. In most (but not all) cases there is a net deposition flux of reducing species, which is balanced by a positive \ce{H2} flux from the ocean to the atmosphere.


Figure \ref{Fig: Figure 2} provides a detail look into the mixing ratio profiles of the key species and shows the difference between the endmember scenarios of the very high emission rate and the very low emission rate. With the very high emission rate, the mixing ratio of O$_2$ drops to extremely small values near the surface for surface pressures ranging from 0.1 to 10 bars. The mixing ratio of CO is $>1\%$ and is larger for lower surface pressures. With the very low emission rate, however, O$_2$ can have substantial mixing ratios at the surface when the atmosphere is larger than $\sim0.3$ bars; the mixing ratio of this \ce{O2} generally increases for larger atmospheres, to about $6\times10^{-4}$ (Figure \ref{Fig: Figure 1}). Due to this O$_2$ accumulation, an ozone layer is formed, with the peak ozone mixing ratio between $10^{-6}$ and $10^{-5}$, on the same orders of magnitude as the present-day Earth's ozone layer.

Pressure-dependent degassing of magma can impact the oxygen buildup when the overall emission rates are small. Figure \ref{Fig: Figure 2x} compares the standard model, where the same volcanic gas composition is assumed for all surface pressures, and the special case, where 10-fold less \ce{H2} and \ce{SO2} in volcanic gas is assumed for the 10 bar surface pressure atmosphere, according to the calculations of \cite{GAILLARD2014307} for degassing under the surface pressure. There is no significant difference in the abundance of atmospheric \ce{O2} for the very high emission scenario, but the \ce{O2} abundance increases by 80\% (to $1.1\times10^{-3}$) when the pressure dependency of degassing is taken into account. The main cause is a further reduction of the \ce{H2} emission rate when the surface pressure is high. The atmospheres with less \ce{H2} emission have less \ce{H2} and more CO and \ce{O2}.

Why are the behaviors of the high emission scenarios and the low emission scenarios so different when changing the surface pressure? Figure \ref{Fig: Figure 3} shows the reaction rate profiles of key chemical reactions that lead to the combination of CO and O. It is well known that the direct combination
\reaction{CO + O + M -> CO2 + M\label{r:1}}
requires a third molecule and is only efficient when the ambient number density is large. The combination can alternatively proceed with
\reaction{CO + OH -> CO2 + H\label{r:2}}
For all modeled scenarios, we find that \ref{r:1} dominates the lower atmosphere and \ref{r:2} dominates the middle and the upper parts of the atmosphere. These two combined can account for the photodissociation of \ce{CO2} (Figure \ref{Fig: Figure 3}). For \ref{r:2} to work as a catalytic cycle, H has to be converted to OH. The following cycle is known to operate in the atmosphere of Mars \citep{Nair1994}:
\reaction{H + O2 + M -> HO2 + M\label{r:3}}
\reaction{O + HO2 -> OH + O2\label{r:4}}
The net result of \ref{r:3} and \ref{r:4} is \ce{H + O -> OH}; and this completes the catalytic cycle of \ref{r:2} that combines CO and O. Figure \ref{Fig: Figure 3} shows that in the very low emission scenario, this cycle indeed operates and dominates the combination in the middle atmosphere. However, in the very high emission scenario, we find that the rate of \ref{r:3} is substantially smaller than the rate of \ref{r:4} in the middle atmosphere. Instead, the following cycle produces the \ce{HO2} needed in \ref{r:4}:
\reaction{H + CO + M -> CHO + M \label{r:5}}
\reaction{CHO + O2 -> CO + HO2 \label{r:6}}
The net result of \ref{r:5} and \ref{r:6} is exactly the same as \ref{r:3}. This cycle replaces \ref{r:3} and dominates the middle atmosphere in the very high emission scenario.

Therefore, the starting reaction for catalytic combination of CO and O is different between the very high emission scenario and the low and very low emission scenarios. For the former it is \ce{H + O2} (\ref{r:3}) and the latter it is \ce{H + CO} (\ref{r:5}). Moreover, the pressure dependency of the reaction rate constants of \ref{r:3} and \ref{r:5} is different. The reaction rate constant of \ref{r:3} has a high-pressure limit of $7.5\times10^{-11}(T/300)^{0.21}$ cm$^3$ s$^{-1}$ where $T$ is the temperature \citep{Burkholder}. The reaction rate constant of \ref{r:5} does not have an established high-pressure limit and is $5.3\times10^{-34}\exp(-370/T)N$ cm$^6$ s$^{-1}$, where $N$ is the ambient number density \citep{Baulch1994}. The rate constant of \ref{r:5} is $3.7\times10^{-15}$ cm$^3$ s$^{-1}$ at 1 bar and 300 K, and is 10 times higher at 10 bar. Comparing the rate constants and the O$_2$ and CO mixing ratios, we find that with the very high emission rate, the mixing ratio of CO is more than four orders of magnitude higher than that of O$_2$ (Figure \ref{Fig: Figure 1}); and so, \ref{r:5} dominates over \ref{r:3}. As the cycle starting with \ref{r:5} becomes more efficient at higher pressures, the mixing ratio of \ce{O2} at the steady state decreases. With the very low emission rates, \ref{r:3} dominates over \ref{r:5}, and its rate constant reaches the high-pressure limit. The catalytic cycle cannot become more efficient as the pressure increases, leading to an increase of the steady-state \ce{O2}.

\section{Discussion} \label{Discussion}

The simulations presented here qualitatively confirm the results of \cite{2012ApJ....761...166H} at the surface pressure of 1 bar and extend to the surface pressures ranging from 0.1 to 10 bars. Compared to \cite{2012ApJ....761...166H}, we have in this work changed the main reducing species in outgassing from \ce{CH4} to \ce{CO} according to the subaerial magma degassing model of \cite{GAILLARD2014307}, and enforced the redox balance of the ocean by including a \ce{H2} return flux controlled by the redox balance of the deposition \citep{Harman2015}. We have also worked with a more realistic range of volcanic activities, in that the ``zero'' emission case in  \cite{2012ApJ....761...166H} is replaced by the ``very low emission'' scenario in this work corresponding to Venus's volcanic rate. With these updates, the results continue to indicate accumulation of \ce{O2} near the surface of the planet. The cases where the \ce{O2} mixing ratio is indefinitely small at the surface are the very high emission scenarios (20 times higher than present-day Earth's volcanic activity) and the 0.1 bar surface pressure scenarios. All other cases have \ce{O2} near the surface to various abundances, and also have \ce{O3} peaks in the middle atmosphere. This result is different from \cite{Harman2015}, where the authors suggest the \ce{O2} accumulation for terrestrial planets of Sun-like stars is completely prevented by enforcing the redox balance of the ocean. The difference may be due to the reaction networks being used, and also the criteria of convergence in the model.

The steady-state mixing ratio of \ce{O2} is below $10^{-3}$, the presumed level for \ce{O2} to be detectable via direct imaging, for all model scenarios. The only exception is the 10-bar atmosphere under the very low volcanic activity level and with the further reduced emission rate of \ce{H2} due to the high surface pressure (i.e., the dashed lines in Figure \ref{Fig: Figure 2x}). Therefore, our model results, while different from \cite{Harman2015} in terms of the mixing ratio profiles of \ce{O2}, support their generic conclusion that \ce{O2} is unlikely to have a photochemical false positive if found in abundance ($>10^{-3}$) on water-rich and terrestrial planets in the habitable zone of a Sun-like star. This finding supports the use of \ce{O2} as a key indicator for potentially habitable worlds by future direct imaging space missions. 

Our models further indicate that the potential for photochemical oxygen to accumulate to nearly the detectable level is particularly large when the atmosphere has a large amount of \ce{CO2} and when the volcanic activity of the planet is low. The planets close to the outer edge of the habitable zone are expected to have large CO$_2$ partial pressures in the atmosphere \citep{Kasting:1993zz}. This risk can be mitigated by detecting signatures of volcanic activities. The signatures include sulfur-bearing species \citep{Kaltenegger2010,Hu2013} and methane \citep{2014ApJ....792...90D}. 
Finally, we note that while the steady-state mixing ratios of \ce{O2} in our fiducial cases are not particularly larger than $10^{-3}$, they are in some cases larger than the oxygen level of Earth in most of the Proterozoic Eon after the Great Oxidation Event \citep{Planavsky2014}. Detail photochemistry models are thus necessary to intepret future detections of \ce{O2}.

In addition to \ce{O2}, our simulations suggest substantial photochemical production of \ce{O3} over a wide range of surface pressures and emission rates. \cite{2014ApJ....792...90D} showed that ozone would be detectable via its Hartley band at 0.25 $\mu$m for a column depth of $10^{15}\sim10^{18}$ cm$^{-2}$. The column depth of \ce{O3} we find from the very low to the high emission scenarios spreads in this range (Figure \ref{Fig: Figure 1X}). Ozone alone cannot be considered as a biosignature gas due to this false positive, in agreement with \citep{2014ApJ....792...90D}. The ozone feature at 0.25 $\mu$m is particularly sensitive to the atmospheric scenarios determined by the emission rates and the surface pressure. A measurement in this band would thus greatly help to pinpoint the atmospheric scenario of a terrestrial exoplanet.


Finally, the trend we find for the low and very low emission scenarios is consistent with the work of \cite{Zahnle2008}, in the context of the Martian atmosphere. Despite the difference in terms of the atmospheric water abundance, we essentially agree with \cite{Zahnle2008} that a larger CO$_2$ atmosphere would be more photochemically unstable, if the surface emission rates are low. We discover the reverse trend under the high emission rates. The high atmospheric CO versus O$_2$ ratio leads to the catalytic cycle initiated by \ce{H + CO} (Section \ref{section:results}) and further stabilizes the CO$_2$ atmosphere at high surface pressures. If this trend also applies to dryer conditions, an atmosphere of early Mars more massive than the present might have been stabilized by strong volcanic outgassing.

\begin{table*}[ht]
\caption{Redox Fluxes (Equivalent H cm$^{-2}$ s$^{-1}$) and Column Depths (cm$^{-2}$) of Modeled Atmospheres }
\centering
\begin{tabular}{lccccc}
\hline \hline
\multirow{ 2}{*}{Scenarios}&\multicolumn{2}{c}{Very High Emission}&\multicolumn{2}{c}{Very Low Emission} \\
& 0.1 bar& 10 bar & 0.1 bar & 10 bar\\
\hline
\textit{Escape}\\
H& -7.38E+8 & -2.95E+8 &-7.92E+6 & -7.24E+5\\
H$_2$& -1.37E+11&-1.38E+11&-1.08E+8&-1.14E+8\\
\textit{Total} & -1.38E+11 & -1.38E+11 & -1.16E+8 & -1.15E+8 \\
\hline
\textit{Outgassing}\\
H\textsubscript{2}& 6.0E+10&6.0E+10&5.0E+7&5.0E+7\\
CO&6.0E+10&6.0E+10&5.0E+7&5.0E+7\\
H\textsubscript{2}S& 1.8E+10&1.80E+10&1.5E+7&1.5E+7\\
\textit{Total} &1.38E+11& 1.38E+11&1.15E+8 &1.15E+8 \\
\hline
\hline
\textit{Dry and Wet Deposition}\\
O$_3$ & - & - & - & 3.25E+10 \\
HO\textsubscript{2}
& - & - & 4.96E+8 & 1.28E+7\\
H\textsubscript{2}O\textsubscript{2} & - & - &4.27E+9&7.04E+8\\
CO&-4.17E+10&-6.12E+10&-2.40E+9&-3.52E+10\\
CH\textsubscript{2}O&-9.25E+7& - &-2.63E+8& -\\
H\textsubscript{2}S&-7.39E+9&-1.73E+10&-5.98E+6&-1.44E+7\\
H\textsubscript{2}SO\textsubscript{4}&6.84E+7&9.33E+7&1.34E+6& -\\
S$_8$ & -1.92E+10 &-&2.17E+6& -\\
\textit{Total} & -6.83E+10 & -7.84E+10 &2.10E+9 & -2.03E+9\\
\hline
\textit{Return H$_2$ Flux}\\
H\textsubscript{2} & 6.83E+10 & 7.84E+10 & -2.09E+9 & 2.03E+9\\
\hline
\textit{Column Depth}\\
\ce{O2} & $6.4\times10^{18}$ & $3.0\times10^{18}$ & $2.5\times10^{19}$ & $4.9\times10^{23}$\\
\ce{O3} & $2.0\times10^{14}$ & $8.0\times10^{14}$ & $3.1\times10^{15}$ & $9.4\times10^{18}$\\
\hline
\label{Tab2}
\end{tabular}
\end{table*}

\section{Conclusion}

We have used a 1D photochemical model to simulate the composition of CO$_2$-dominated atmospheres on terrestrial exoplanets in the habitable zone of a Sun-like star, for the surface pressures ranging from 0.1 to 10 bars and the emission rates corresponding to volcanic activities from Venus-like to 20 times higher than Earth-like. Our models maintain the redox balance of both the atmosphere and the ocean. We find that the emission rates control how the mixing ratio of photochemically produced O$_2$ changes with the surface pressure. The mixing ratio of O$_2$ increases with the surface pressure when the emission rates are very low, consistent with previous studies. However, driven by a catalytic cycle initiated by the combination reaction between H and CO, the mixing ratio of O$_2$ decreases with the surface pressure when the emission rates are high. We have also studied the effect of the surface pressure on the speciation of magma degassing and the composition of volcanic outgassing. For the very low volcanic activity and 10-bar atmosphere, this effect almost doubles the steady-state mixing ratio of \ce{O2}.

To search for potentially habitable exoplanets, our models support the use of \ce{O2} detectable via its A band at 0.76 $\mu$m as a key indicator for oxygenic photosynthesis, on terrestrial planets in the habitable zone of Sun-like stars. The maximum amount of photochemical \ce{O2} we find is $1.1\times10^{-3}$ in terms of volumetric mixing ratio. We could define the amount of \ce{O2} above this level to be ``abundant'', and such a definition is natural and commensurate with the strength of the A band and reasonable prospect for detection capabilities in the next one or two decades. Abundant oxygen, with contextual information including the rocky nature of the planet and the existence of water vapor in the atmosphere, is then probably the signature we should aim for. We may miss some habitable planets in this way, since Earth managed to maintain its oxygen level below $10^{-3}$ for much of the time after the rise of oxygenic photosynthesis \citep{Reinhard2017}. We likely have to accept this ``false negative'', as the models presented here show that oxygen lower than $10^{-3}$ has a direct photochemical false positive. This photochemical false positive may however be mitigated by also ruling out abundant CO in the atmosphere \citep[e.g.,][]{Schwieterman2016}, because all our models have large mixing ratios of CO. In all, with the volcanic activity and the pressure dependency of the reaction rate constants quite universal for planets, our numerical experiments provide a useful baseline to understand under what conditions oxygen on potentially habitable terrestrial exoplanets can be regarded as a biosignature.

\acknowledgments
TJ thanks the support of the US National Science Foundation's (NSF) Division Of Undergraduate Education (DUE), under Grant No.1457943.

\bibliographystyle{yahapj}
\bibliography{references}

\end{document}